# Selective resolution of phonon modes in STM-IETS on clean and oxygen-adsorbed Cu(100) surfaces

Minjun Lee[1], Myungchul Oh[1], Hoyeon Jeon[1], Sunwouk Yi[1], Inhae Zoh[1], Chao Zhang[1], Jungseok Chae[2,3*], and Young Kuk[1*]

[1]Department of Physics and Astronomy, Seoul National University, Seoul 08826, Korea
[2]Center for Quantum Nanoscience, Institute for Basic Science, Seoul 03760, Korea
[3]Department of Physics, Ewha Womans University, Seoul 03760, Korea

The observation of surface phonon dispersion using local probes can provide important information related to local structural and thermal properties. In this study, surface phonon modes on a Cu(100) surface were measured using the inelastic tunneling spectroscopy of scanning tunneling microscopy (STM-IETS) with atomically sharp tips. Different phonon modes were selectively measured depending on the structures of the probing tips or the surfaces. Two different surface phonon modes, at 19.0 meV on a clean Cu(100) surface and at 13.5 meV on an oxygen-adsorbed Cu(100) surface, are explained by the selection rules. Additionally, the spatial variation in STM-IETS showed surface stress relaxation.

*To whom correspondence should be addressed: chae.jungseok@qns.science, ykuk@phya.snu.ac.kr

Surface phonon modes can be different from bulk phonon modes because they arise from the abrupt termination of a crystal structure at a surface. Surface phonon dispersion has been measured by electron energy loss spectroscopy (EELS) [1] and He atom scattering (HAS) [2], but these tools measure the averaged dispersions over the beam spot size, limiting the spatial resolution[3]. To understand the local properties of surface phonon modes related to surface structures, a local probe method is required. Since vibration spectroscopy with a local probe was first realized using the inelastic tunneling spectroscopy (IETS) of scanning tunneling microscopy (STM)[4], it has become a powerful tool to measure the molecular vibrational modes on single molecules[5–7] and spin excitation on single magnetic atoms[8-11] on surfaces. Unlike molecular vibrations, which are localized within a single molecule, surface phonons originate from the quantization of collective lattice vibration over the surface. However, despite this difference, STM-IETS may measure the excitation of surface phonon modes with the same measurement principle as that for molecular vibrations. On graphite and graphene surfaces, which have a covalent bonding nature, surface phonon modes were measured using STM-IETS[12,13] at the van Hove singularities (VHS), which are expected from calculation. On metal surfaces, however, acoustic phonon peak was measured with a high peak width, and there was no correlation with VHS[14,15] and higher energy peaks were hard to detect in STM-IETS measurements.

Although the mechanism of opening of an inelastic tunneling channel by a surface phonon is not yet clearly understood, it was suggested that a tunneling electron is coupled to a surface phonon[16]. As a result, the inelastic tunneling signal that we measured is proportional to the product of the electron-phonon coupling matrix elements and the phonon density of states. Consequently, the momentum-average of the tunneling matrix limits the energy resolution of the phonon mode compared with the molecular vibration spectra in STM-IETS. Although the physical reasons for opening an inelastic tunneling channel are different between the cases of surface phonons and molecular vibration, it has been reported that the size and atomic configuration of the tip determines the STM-IETS signal strength[17]. A single-atom tip yields an enhanced STM-IETS signal at a CO molecule on the Cu(111) surface compared with three- or four-atom tip configurations. According to theoretical studies on STM-IETS[18], the STM-IETS signal is dependent on the relaxation energy of the atom under the tip, which opens the inelastic tunneling channels and can be affected by the presence of a sharp STM tip. Based on this experimental and theoretical background, we measured the surface phonon modes on a clean Cu(100) surface and oxygen-adsorbed Cu(100) surfaces by using atomically sharp STM tips prepared by field ion beam (FIB) sharpening (Fig. S1 in the supplemental material), followed by *in situ* sputtering and annealing. Several acoustic phonon modes were measured on these surfaces by using these sharp tips. Moreover, a spatial variation map on oxygen-adsorbed surfaces shows locally enhanced STM-IETS signals, possibly due to local surface relaxation.

Before exploring oxygen-adsorbed surfaces, a clean Cu(100) surface was prepared by several cycles of Ar sputtering and annealing at 450 °C. Figure 1(a) shows an STM topographic image of the clean Cu(100) surface, and STM-IETS spectra were acquired at the clean surface as a reference to observe the change in phonon mode peaks after oxygen exposure. All STM measurements in this paper were performed at a temperature of ~5 K. Figure 1(b) shows a representative STM-IETS spectrum that reveals phonon modes on the clean Cu(100) surface at 3.7, 19.0, and 27.9 meV, which were extracted by Lorentzian fitting. Hereafter, these peaks will be referred to as V1, V3, and H1, respectively. The lowest-energy phonon mode at 3.7 meV was measurable in previous STM-IETS studies on metal surfaces. The observation of this peak is under debate because no VHS exists at this energy in the phonon density of states. Vitali *et al.*[12] suggested that the lowest-energy mode could have originated from the atomic vibration at the tip apex. However, E. Minamitani *et. al.*[15] proposed that the out-of-plane polarized surface phonons near the Γ point can be measured as a peak in STM-IETS. In our experiment, the lowest-energy peak was generally measurable with all different STM tips and samples. The higher-energy peaks at 19.0 meV and 27.9 meV can be assigned as out-of-plane and in-plane phonon modes, respectively, which were previously measured in an EELS experiment[19,20] and confirmed by density functional theory[21,22]. To our knowledge, the phonon mode peak at 19.0 meV has been resolved for the first time using STM-IETS. We believe that a geometrically sharper STM tip increases the relaxation energy by forming an electrostatic potential profile with a higher spatial gradient around the atom to increase the STM-IETS signal.

After measuring clean Cu(100), the surface was exposed to oxygen to cause relaxation on the surface. To grow the ordered oxygen-adsorbed layer, a clean Cu surface was heated to 450 °C at $1 \times 10^{-10}$ Torr followed by exposure to oxygen at ~3000 Langmuir at 200 °C [23]. Figure 2(a) shows an STM topographic image of the oxygen-adsorbed Cu(100) surface, in which two different surface reconstruction regions coexist. One region appears flat, while the other appears rough. The line profile across these two regions clearly shows different roughness values of approximately 6 pm and 25 pm, respectively, as shown in Fig. 2(b). Based on previous studies[23,24] and the magnified STM image (Fig. 2(c)), the atomic structure of the flat region is revealed to have (2√2 × √2)R45°-O surface reconstruction and that of the rough region is revealed to have (√2 × √2)R45°-O surface reconstruction. In the (2√2 × √2)R45°-O surface, oxygen atoms are located at the center hollow sites of the Cu lattice with consecutive missing Cu rows, as shown in Fig. 2(d). The missing rows are clearly visible at the center rectangular region of the magnified STM image (Fig. 2(c)), as indicated by black arrows in Fig. 2(d). In the (√2 × √2)R45°-O surface, oxygen atoms are located at the same sites as in the (2√2 × √2)R45°-O surface, but no rows with missing Cu exist, as shown in Fig. 2(e).

To clarify the factors determining the STM-IETS resolution, we measured the STM-IETS spectra on the (2√2 × √2)R45°-O surface with different modulation voltages. Figure S2 (a) in the supplementary material shows STM-IETS spectra obtained with different modulation voltages of 3, 5, 10, and 20 mV. Features that could be related to atomic vibrations at the tip were not observed in our measurements. Instead, the phonon modes at energies of 3.7 meV (V1) and 13.5 meV (V2) were clearly resolved with a modulation voltage of 3 mV (tip #1), and peak energies of 3.7 meV (V1) and 27.9 meV (H1) were resolved with a modulation voltage of 5 mV (tip #2). This discrepancy may have originated from the selective sensitivity of the local geometry of the tip to specific phonon modes. In this case, tip #1 is sensitive to out-of-plane phone modes, and tip #2 is sensitive to in-plane modes. However, the microscopic origin of this difference is not clear yet. With larger modulation voltages of 10 and 20 mV, the lowest-energy peak is shifted to larger energies of 7.7 meV and 13.0 meV, respectively, because of the convolution of two or, possibly, three peaks.

To study the effect of oxygen atoms on surface phonon modes, STM-IETS spectra obtained on the (2√2 × √2)R45°-O surface were compared with that obtained on the clean Cu(100) surface, as shown in Fig. 3(a). First, the lowest-energy mode (V1) is measured at the same energy of 3.7 meV in all three surfaces (the clean and two oxygen-adsorbed surfaces), implying that oxygen atoms do not affect the out-of-plane polarized surface phonon mode near the Γ point. Further, this mode remains at the same energy with different tips. Second, the out-of-plane surface mode at 13.5 meV (V2) on the (2√2 × √2)R45°-O surface, which was previously resolved by EELS measurement[20], was measured for the first time using STM-IETS. This peak was not resolved on the clean Cu(100) surface, although it should exist at the same energy. The resolution of this peak can be explained by the symmetry of the phonon mode. The V2 peak is related to the out-of-plane surface mode peak at the X point of the Brillion zone boundary in the phonon dispersion relation [22], which is, owing to symmetry, parallel to the direction of the Cu-missing rows in real space, as shown in Fig. 3(b). Then, phonons can be strongly localized along the edges near the Cu-missing rows, as reported by L. Niu *et al.*[25], which causes the enhancement in the STM-IETS signal in combination with the symmetry of STM tip #1. Interestingly, another out-of-plane surface mode at 19.0 meV (V3), which was resolved on the clean Cu(100) surface, was suppressed on the $(2\sqrt{2} \times \sqrt{2})R45°$-O surface. In contrast to the V2 peak, the V3 peak is related to the surface phonon mode at the M point of the Brillion zone boundary in the phonon dispersion relation[22], which is perpendicular to the direction of Cu-missing rows in real space, as shown in Fig 3(c). In the case of peaks V2 and V3, the symmetry of the phonon modes enables the selective resolution of peaks in the STM-IETS measurement. Moreover, the in-plane mode measured at 26.5 meV (H1) on the clean Cu(100) surface was shifted to 25.5 meV on the $(2\sqrt{2} \times \sqrt{2})R45°$-O surface, although the difference is within the range of our error bar. This result is consistent with a previous result obtained using low-energy electron diffraction[26], which was explained by the compressive stress to Cu atoms on the surface induced by oxygen atoms. On the $(\sqrt{2} \times \sqrt{2})R45°$-O surface, as shown in Fig. S2(b), only the V1 peak is clearly visible, and the remaining peaks are all suppressed because the random occupation of oxygen atoms makes the STM tip remain far away from the surface under the same measurement conditions, weakening the STM-IETS signal.

To investigate local variation in phonon modes, $d^2I/dV^2$ maps were obtained over a coexisting area between two oxidized reconstruction surfaces. Figures 4(a) and (b) show an STM topographic image and a $d^2I/dV^2$ map of the same area focused on the V2 peak. The topographic image shows the (2√2 × √2)R45°-O phase at the center and the (√2 × √2)R45°-O phase on the right (same layer as the center phase) and left (one Cu layer lower than the center phase). The V2 peak position is not changed over the whole area, as described previously, but interestingly, the STM-IETS signal intensity varies locally in the $d^2I/dV^2$ map (Fig. 4(b)), as indicated by the bright lines from the bottom left to top right. It is noted that, with a bias voltage of -15 mV for the map, the brighter part represents a stronger $d^2I/dV^2$ signal related to the V2 phonon mode. One possible reason is that the surface stress at an oxygen-adsorbed surfaces is changed from the bare Cu surface, and the stress is relaxed to form narrow line-shaped patterns in a certain direction with a length scale of ~1-2 nm, as shown in Fig. 4(c). A similar stress relaxation was observed on the double Fe layer on the W(100) surface[27,28]. The most stressed narrow region can experience different local electrostatic potential profiles, which easily open the inelastic channel in STM-IETS. For example, the increase in the STM-IETS signal was reported along the boundary between the face-centered-cubic and hexagonal-closed-packed region on an Au(111) surface[14]. The enhancement in the phonon mode was measured on graphene nano-bubbles owing to the geometrical structure[29]. These results can be explained by the stress relaxation due to the boundary or structural deformation and the increase in the STM-IETS signal. For further understanding, additional experimental and theoretical works are needed.

In conclusion, surface phonon modes on clean and oxygen-adsorbed Cu(100) surfaces were successfully resolved using the STM-IETS technique. A geometrically sharp STM tip is an important factor to resolve many phonon modes on the surface. Additionally, the symmetry of the phonon modes, in combination with the STM tip symmetry, is directly related to the STM-IETS signal strength. The spatial variation of the STM-IETS signal shows the possibility that broken symmetry due to surface stress relaxation can be another factor determining the STM-IETS signal.

In the future, we plan to work on STM-IETS measurements with atomic resolution near the Cu-missing rows or the stressed strips for further understanding of the microscopic origin of the STM-IETS signal on the surface.

Acknowledgements

This work is supported in part by the National Research Foundation of Korea (NRF) Grant (NRF-2006-0093847, NRF-2010-00349). JSC was supported by the IBS research program IBS-R027-D1.

**Figure Captions:**

Figure 1. (a) STM topographic image of a clean Cu(100) surface obtained in the constant current mode with a tunneling current of 10 pA and sample bias of 2 V (scale bar: 30 nm). (b) Representative STM-IETS spectrum obtained on a clean Cu(100) surface with a modulation voltage of 3.3 mV. The spectrum is fitted by three Lorentz curves centered at 3.7, 19.0, and 27.9 mV, respectively.

Figure 2. (a) STM topographic image of the copper-oxide surface obtained with a tunneling current of 50 pA and sample bias of 2 V (scale bar: 50 nm). (b) Height profile taken along the blue line in (a). (c) High-resolution image of the copper-oxide surface obtained with a tunneling current of 100 pA and sample bias of -50 mV. Cu missing rows in the $(2\sqrt{2} \times \sqrt{2})R45°$-O region are clearly visible, as explained in the main text (scale bar: 5 nm). (d) Schematic of atomic arrangement on the $(2\sqrt{2} \times \sqrt{2})R45°$-O surface. The black rectangle shows a unit cell of $(2\sqrt{2} \times \sqrt{2})R45°$-O, and black arrows indicate Cu missing rows. (e) Schematic of atomic arrangement on the $(\sqrt{2} \times \sqrt{2})R45°$-O surface. The black rectangle shows a unit cell of the $(\sqrt{2} \times \sqrt{2})R45°$-O surface, and oxygen atoms randomly occupy the center hollow sites.

Figure 3. (a) STM-IETS spectra obtained on the $(2\sqrt{2} \times \sqrt{2})R45°$-O surface (blue and red) with different STM tips and on the bare Cu(100) surface (black). Arrows indicate the energy positions of phonon modes with V2 (red), V3 (black), and H1 (blue), respectively. (b, c)

Schematics of the real space direction of V2 (red arrow in b) and V3 (black arrow in c) phonon modes, respectively. (d) Energy of the phonon modes on the bare Cu and (2√2 × √2)R45°-O surfaces.

Figure 4. (a) STM topographic image of a copper-oxide surface obtained with a tunneling current of 100 pA and sample bias of 500 mV. (b) STM-IETS map of the same area as in (a) with a tunneling current of 100 pA and bias voltage of -15 mV. All scale bars indicate 5 nm. (c) Height profile across the black line in (b).

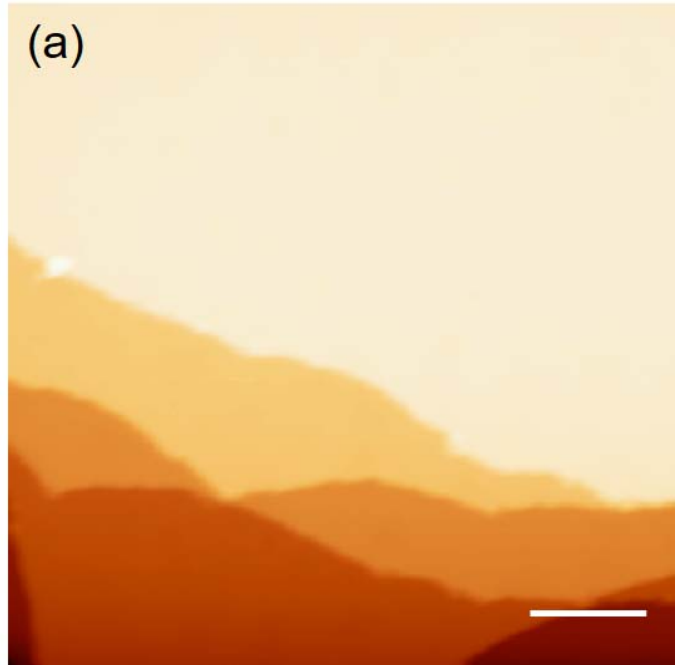

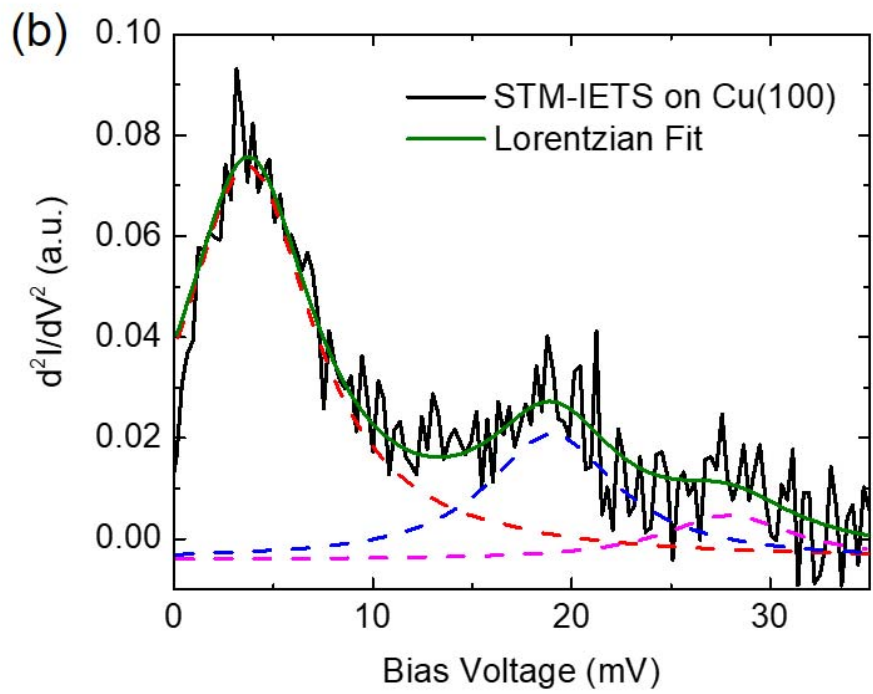

Fig. 1: Lee

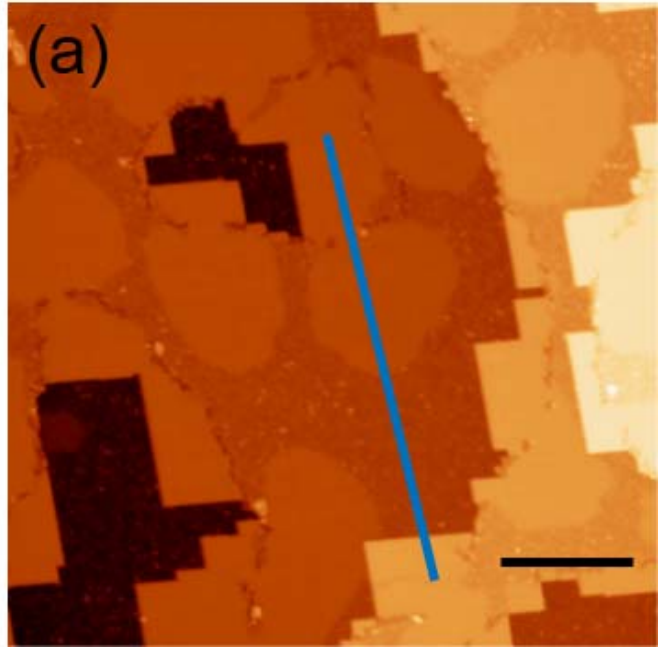

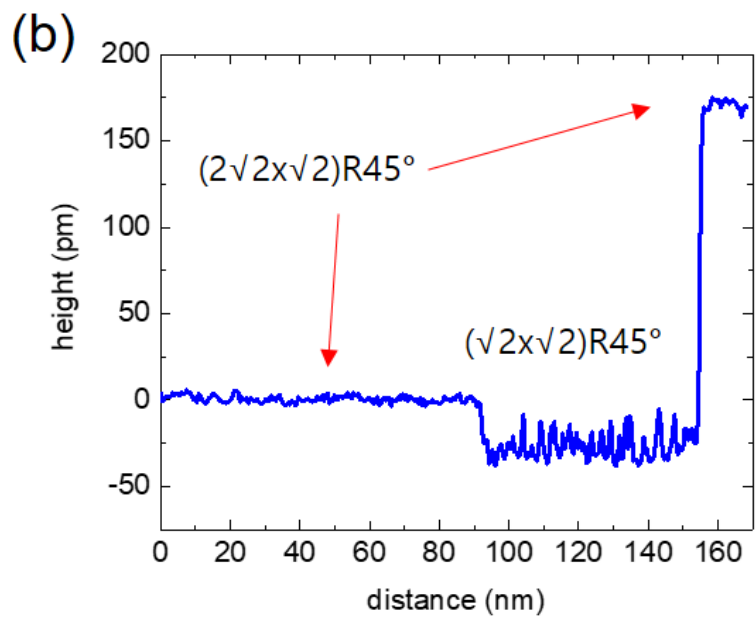

Fig. 2: Lee

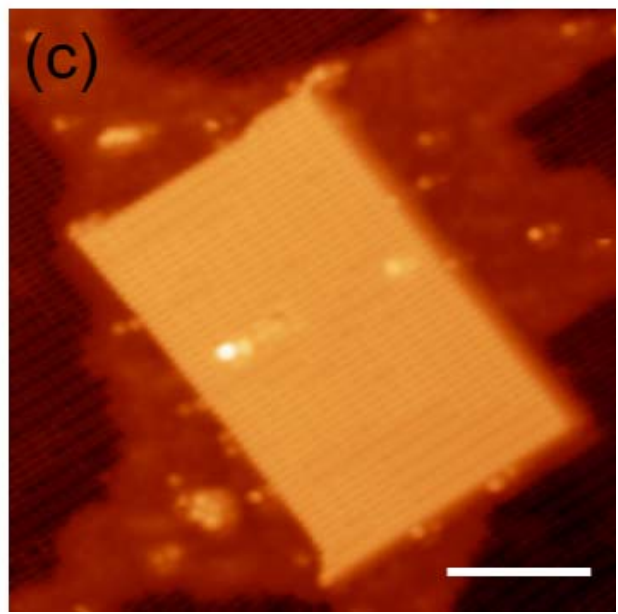

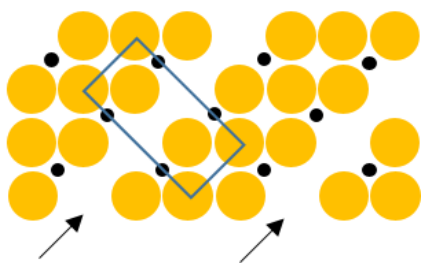

(d) (2√2x√2)R45°

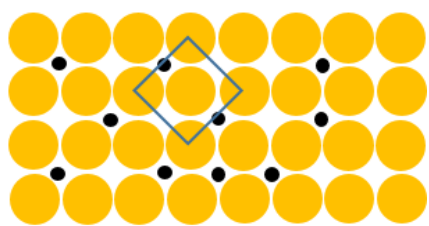

(e) (√2x√2)R45°

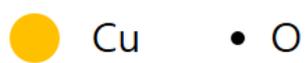

● Cu    • O

Fig. 2: Lee

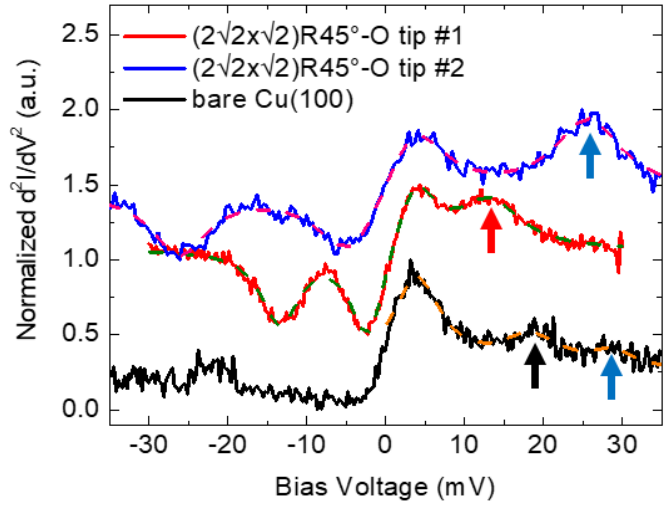

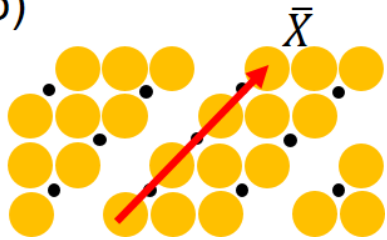
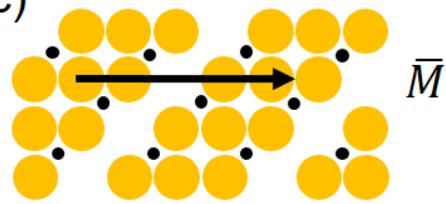

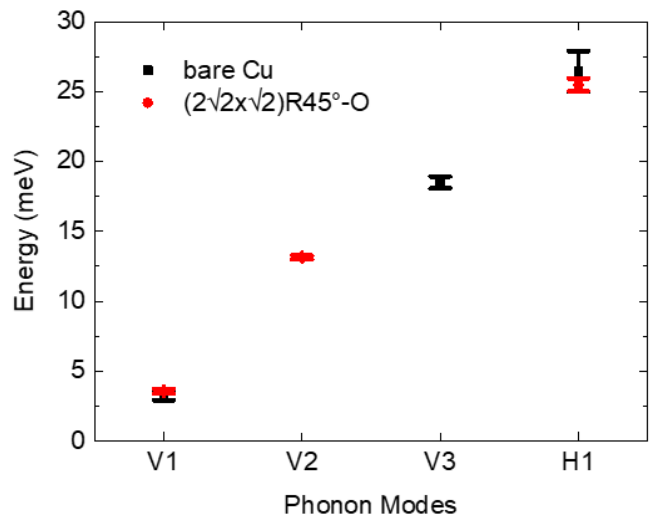

Fig. 3: Lee

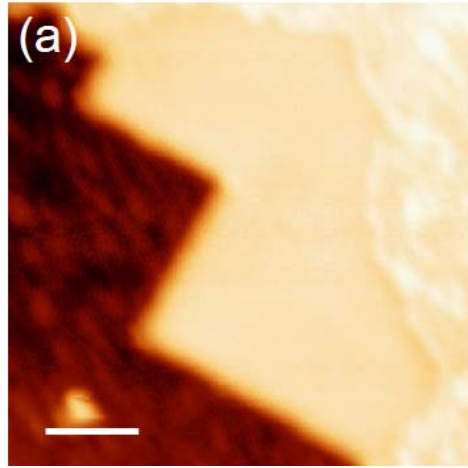

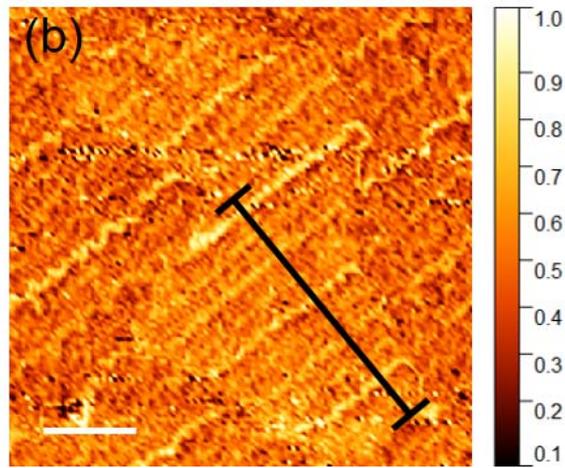

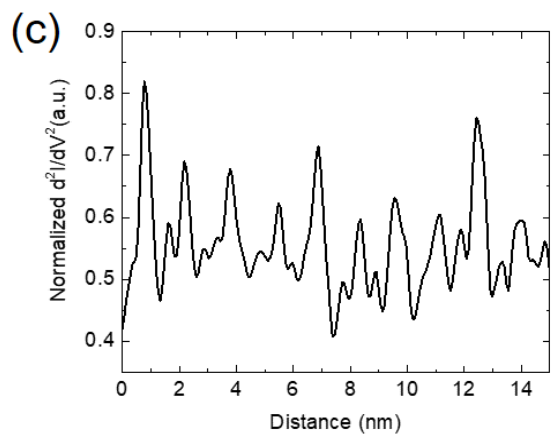

Fig. 4: Lee

# Selective resolution of phonon modes in STM-IETS on clean and oxygen-adsorbed Cu(100) surfaces

Minjun Lee[1], Myungchul Oh[1], Hoyeon Jeon[1], Sunwouk Yi[1], Inhae Zoh[1], Chao Zhang[1], Jungseok Chae[2,3*], and Young Kuk[1*]

[1]Department of Physics and Astronomy, Seoul National University, Seoul 08826, Korea
[2]Center for Quantum Nanoscience, Institute for Basic Science, Seoul 03760, Korea
[3]Department of Physics, Ewha Womans University, Seoul 03760, Korea

## Table of Contents

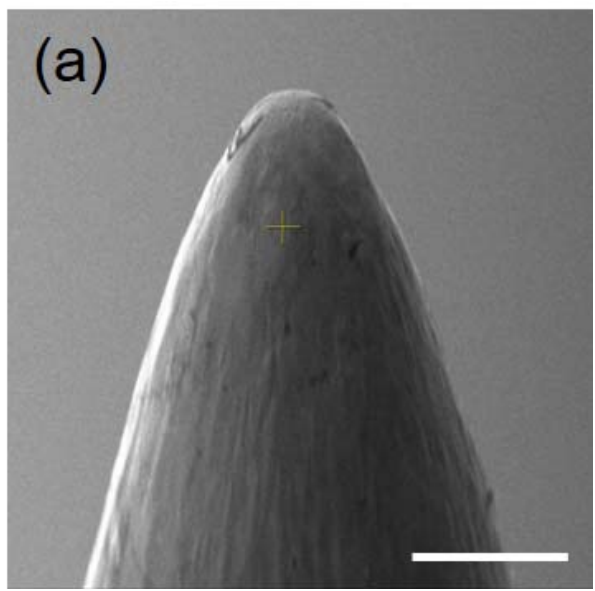

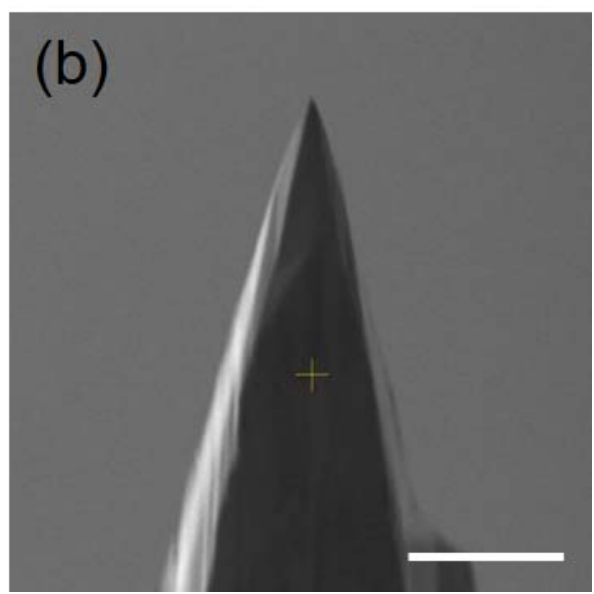

Figure S1. (a,b) SEM image of Ir tip before (a) and after (b) FIB sharpening. (Scale bars: 10 μm)

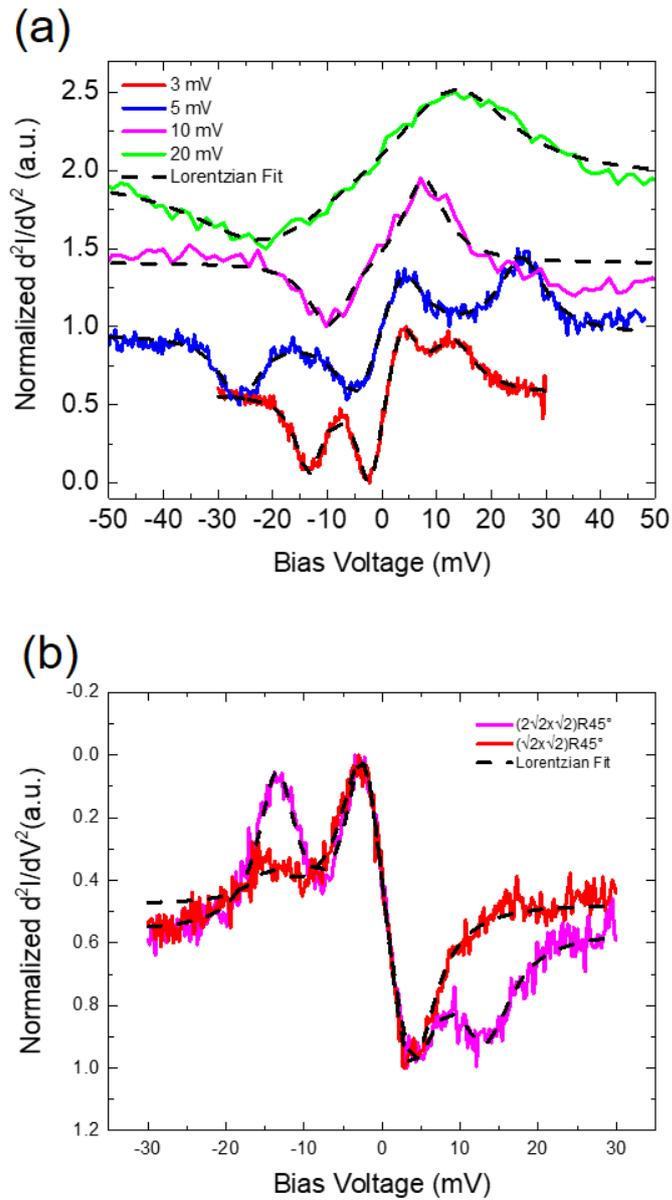

Figure S2 (a) Modulation voltage dependence of IETS peaks with 3 mV (red), 5 mV (blue), 10 mV (pink), and 20 mV (green). (b) STM-IETS spectra for (2√2x√2)R45°-O and (√2x√2)R45°-O surface.